\documentclass[12pt]{article}
\usepackage{graphicx}
\usepackage{amsmath}
\usepackage{amsfonts}
\usepackage{amssymb}    
\usepackage{graphicx}   
\newcommand{\be}{\begin{equation}}
\newcommand{\ee}{\end{equation}}
\newcommand{\bea}{\begin{eqnarray}}
\newcommand{\eea}{\end{eqnarray}}
\newcommand{\ba}{\begin{array}}
\newcommand{\ea}{\end{array}}

\newcommand{\eq}[1]{eq.~(\ref{eq:#1})}

\newcommand{\no}{\nonumber}
\newcommand{\ff}{f}
\newcommand{\cO}{{\cal O}}
\newcommand{\cA}{{\cal A}}

\begin{document}

\begin{flushright}
July 2005 \\
\end{flushright}
\vskip   1 true cm 
\begin{center}
{\Large \textbf{Electromagnetic corrections \\ [2 pt]
to non-leptonic two-body $B$ and $D$ decays}}    \\ [20 pt]
\textsc{Elisabetta Baracchini}${}^{1}$ and \textsc{Gino Isidori}${}^{2}$   \\ [20 pt]
${}^{1}~$\textsl{ Dipartimento di Fisica, Universit\'a di Roma
``La Sapienza'' and \\ INFN, Sezione di Roma, P.le A. Moro 2,
I-00185 Roma, Italy  } \\ [5pt]
${}^{2}~$\textsl{INFN, Laboratori Nazionali di Frascati, I-00044 Frascati,
      Italy} 

\vskip   1 true cm 

\textbf{Abstract}
\end{center}
\noindent  
We present analytic expressions to evaluate at $\cO(\alpha)$ 
the effects of soft-photon emission, and the related virtual corrections, 
in non-leptonic decays of the type 
$B,D \to P_1 P_2$, where $P_{1,2}$  are scalar or pseudoscalar
particles.  The phenomenological implications 
of these results are briefly discussed. 
For $B$ decays into charged pions the effects of soft-photon 
emission are quite large: the corrections to the 
rates can easily exceed the $5\%$ level if tight cuts 
on the photon energy are applied.

\vskip   1 true cm 
          
\section{Introduction}

The large amount of data collected at $B$ factories 
has allowed to reach statistical accuracies of the order 
of a few percent on the measurements of several $B$ meson 
branching fractions. At this level of accuracy
electromagnetic effects cannot be neglected.
On the one hand, in order to ensure a good control of the 
experimental efficiencies, it is necessary to include 
in Montecarlo simulations the unavoidable emission of 
soft photons that accompanies all processes with charged 
particles. On the other hand, the effective cuts applied on 
the (soft) photon spectra are a key information for a 
meaningful comparison between theory and experiments.

The theoretical treatment of the infrared singularities 
generated within QED is a well known subject and one 
of the pillars of quantum field theory. A clear and very 
general discussion can be found, for instance, 
in the classical papers \cite{Yen,Wei}. 
The general properties of QED have been exploited in 
great detail in the case of genuine electroweak 
processes, or processes which can be fully described within 
perturbation theory within the Standard Model (SM).
In these cases there exist both precise theoretical 
calculations of the electromagnetic (e.m.) corrections 
and accurate Montecarlo programs which include the effects 
of soft photon emission, such as PHOTOS \cite{PHOTOS}.
Similar tools have not been developed for most 
exclusive hadronic processes and, in particular, 
for $B$ and $D$ decays. 

Recently, the issue of electromagnetic corrections 
have received considerable attention in the 
context of kaon physics  \cite{Cir,Cir_Kl3,Andre,Gatti}.
As discussed in these works, and as confirmed by 
recent experimental analyses \cite{K-exp}, 
a correct simulation of electromagnetic 
corrections is a key ingredient for a precise 
determination of $V_{us}$ and other
effective couplings of weak interactions.

The purpose of the present work is to present 
simple analytic formulae for he theoretical evaluation,
and the numerical simulation, of the leading radiative 
corrections in $B$ or $D$ meson decays into two
scalar or pseudoscalar particles. Given the universal 
character of the infrared singularities, 
we evaluate the effects of soft photon emission
(and the corresponding virtual corrections) 
within scalar QED and  in the approximation of a point-like 
effective weak vertex. The results thus obtained are valid 
up to constant $\cO(\alpha)$ terms (not enhanced by 
large logs) related to the matching between this 
effective theory and the ``true'' theory where 
the dynamical aspects of weak interactions 
are taken into account. Moreover, our calculation 
does not take into account the possible $\cO(E_\gamma)$ terms 
associated to non-bremsstrahlung amplitudes 
(hard photon emission).

\section{Photon-inclusive widths}

The most convenient infrared-safe observable related to the process
$H \to P_1 P_2$ is the photon-inclusive width
\be
\Gamma^{\rm incl}_{12}(E^{\rm max})  = \left. \Gamma (H \to P_1 P_2 + n\gamma)\, 
\right|_{\sum E_{\gamma} < E^{\rm max} }~,
\ee
namely the width for the process $H \to P_1 P_2$ accompanied by any number 
of (undetected) photons, with total missing energy less or equal to 
$E^{\rm max}$  in the $H$ meson rest frame.
At any order in perturbation theory we can decompose $\Gamma^{\rm incl}_{12}$ 
in terms of two theoretical quantities: the so-called non-radiative width, 
$\Gamma^0_{12}$, and the corresponding energy-dependent e.m.~correction 
factor $G_{12}(E^{\rm max})$,
\be
\Gamma^{\rm incl}_{12}(E^{\rm max}) = \Gamma^0_{12} ~ G_{12}(E^{\rm max})~.
\label{eq:prod}
\ee
The energy dependence of $G_{12}(E)$ is unambiguous and 
universal (i.e.~independent from the short-distance dynamics 
which originate the decay) up to terms which vanish in the 
limit $E \to 0$. On the contrary, the normalization of 
$G_{12}(E)$ is arbitrary: we can always
move part of the finite (energy-independent)
electromagnetic corrections from $\Gamma^0_{12}$
to $G_{12}(E)$. Only the product in (\ref{eq:prod})
corresponds to an observable quantity. 

In the following we report the explicit expressions 
of $G_{12}(E)$  as obtained by means of 
a scalar-QED calculation at $\cO(\alpha)$.
In order to treat separately infrared (IR) and ultraviolet (UV)
divergences, we regulate the former by means of a photon mass 
and the latter by means of dimensional regularization.
We then renormalize the point-like weak coupling 
in the $\overline{\rm MS}$ scheme. 

By construction, we define the non-radiative amplitude $\Gamma^0_{12}$ as follows
\bea
\Gamma^0_{12} &=& \frac{\beta}{16 \pi M_H} \left| \cA_{H \to P_1 P_2}(\mu) \right|^2~,
\label{eq:Gamma12} \\
\beta^2 &=& \left[ 1- \left(r_1+r_2\right)^2\right] \left[ 1- \left(r_1-r_2\right)^2\right]~,
\qquad r_{i} ~=~ \frac{M_i}{M_H}~,
\eea
namely the tree-level rate expressed in terms of the renormalized weak coupling. 
With this convention, the function $G_{12} (E)$ can be written as
\be
G_{12} (E) = 1 + \frac{\alpha}{\pi} \left[ b_{12} \ln\left(\frac{ M_H^2 }{4E^2} \right)
+ F_{12} + \frac{1}{2} H_{12} + N_{12}(\mu) \right]~,
\label{eq:G12}
\ee
where, following the notation of Ref.~\cite{Cir}, we have 
denoted by $H_{12}$ the finite term arising from virtual corrections,
and by $F_{12}$ the energy-independent term generated by the real emission:
\be
\int_{E_\gamma < E }  ~\frac{d^3 \vec{k} }{(2\pi)^3 ~2 E_\gamma}~ 
\sum_{\rm spins} \left| \frac{\cA (H \to P_1 P_2\gamma)}{ \cA  (H \to P_1 P_2)} \right|^2
=~ \frac{\alpha}{\pi} \left[ b_{12} \ln\left(\frac{ m^2_\gamma }{4E^2 } \right)
+  F_{12} +\cO\left( \frac{E}{M_H} \right)\,  \right]~.
\ee
As expected, after summing real and virtual corrections the infrared logarithmic 
divergence cancel out in $G_{12}(E)$, giving rise to the universal 
$\ln(M_H/E)$ terms proportional to
\bea
b_{+-} &=& \frac{1}{2} - \frac{ 4-\Delta_1^2-\Delta_2^2 +2\beta^2 }{8\beta}  \,
  \ln  \left( {\frac {\Delta_1+\beta}{\Delta_1-\beta}} \right) 
\quad  + \quad (1 \to  2)~, \no \\  
b_{\pm 0} &=& 1 - \frac{\Delta_1}{2\beta} \,\ln  
\left( {\frac {\Delta_1+\beta}{\Delta_1-\beta}} \right)~, 
\label{eq:b12} 
\eea
where $\Delta_{1(2)}  = 1 + r_{1(2)}^2 - r_{2(1)}^2$.

Note that $G_{12} (E)$ does depend explicitly 
on the ultraviolet renormalization scale $\mu$. 
The scale dependence contained in $N_{12}(\mu)$ cancels 
out only in the product $\Gamma^0_{12} ~ G_{12}(E)$
due to the corresponding scale dependence of the weak coupling. 
As we shall discuss 
in more detail later on, in practice we are not able to 
exploit the numerical consequences of this cancellation
due to the absence of a first-principle calculation of 
$\cA_{H \to P_1 P_2}(\mu)$.

In the $H^0 \to P_1^+ P^-_2$ case we find the following 
explicit expressions for the coefficients in \eq{G12}:
\bea
F_{+-} &=& \frac{ \Delta_1}{  2\beta  }
          \ln \left( \frac{\Delta_1+ \beta}{ \Delta_1- \beta} \right)  
+ \frac{ 4-\Delta_1^2-\Delta_2^2 +2\beta^2 }{4\beta} 
\left[ \,\ff \left( -\frac{\beta}{\Delta_1} \right) 
- \,\ff \left( \frac{\beta}{\Delta_1} \right) 
\right. \no\\ && 
-\frac{1}{2}\,\ff \left(  \frac{\Delta_1-\beta}{ 2\Delta_1} \right) 
+\frac{1}{2}\,\ff \left(  \frac{\Delta_1+\beta}{ 2\Delta_1} \right)  
+\frac{1}{2}\,\ln  \left( 2 \right) \ln  \left( \frac{\Delta_1-\beta}{\Delta_1+\beta} \right)
 \no\\ && \left.
+\frac{1}{4}\,  \ln^2 \left( 1+\frac{\beta}{\Delta_1} \right) 
-\frac{1}{4}\,  \ln^2 \left( 1-\frac{\beta}{\Delta_1} \right)  
 \right]
  \quad + \quad (1 \to  2)
\eea
\bea
H_{+-} &=& \frac{ 4-\Delta_1^2-\Delta_2^2 +2\beta^2 }{8\beta}  
\left[ {\pi }^{2}+ 2 \ff \left( \frac{\Delta_1+\beta}{2\beta} \right) 
-2 \ff \left( -\frac{\Delta_1-\beta}{2\beta} \right) 
\right. \no\\ && \left. 
+ \ln^2\left(\Delta_1-\beta \right)  -  \ln^2  \left( \Delta_1+\beta \right)  
+2 \ln  \left( {\frac {\Delta_1-\beta}{\Delta_1+\beta}} \right) \ln  \left( \frac{\beta}{2} \right)  \right]
-2\,\ln  \left( 2 \right)  \no\\ && 
+1+\frac{1}{2}\,\beta\,\ln  \left( {\frac {\Delta_1+\beta}{\Delta_1-\beta}} \right) 
 +\frac{1}{2}\, \left(1+ \Delta_1 \right) \ln  \left( {\Delta_1}^{2}-{\beta}^{2} \right) 
\quad + \quad (1 \to  2) \quad   \\ && \no \\
N_{+-} &=& \frac{3}{4} \ln\left(\frac{\mu^2}{M^2_H}\right) 
- \frac{3}{4} \ln(r_1^2)  \quad + \quad (1 \to  2)~,
\eea
where 
\be
f(x) = {\rm Re}[{\rm Li}_2(x)] = - \int_0^x \frac{dt}{t} \ln|1-t|~.
\ee
In the  $H^\pm \to P_1^\pm P^0_2$ case we find:
\bea
F_{\pm 0} &=& 1+\, \frac{ \Delta_1 }{ 2 \beta }
\ln  \left( {\frac {\Delta_1+\beta}{\Delta_1-\beta}} \right)  
-\frac{\Delta_1}{4\beta}\, 
\left[  \ln^2  \left( {\frac {\Delta_1-\beta}{\Delta_1+\beta}} \right) 
+4\,\ff \left( {\frac {2 \Delta_1}{\Delta_1+\beta}} \right)  \right] \\ && \no \\
H_{\pm 0} &=&  -\frac{\Delta_1}{\beta} 
\left[ \frac{1}{2} \ln^2 \left( \Delta_1+\beta \right) 
 -\frac{1}{2} \ln^2 \left( \Delta_1-\beta \right)  
+\ff \left( \frac{\Delta_2+\beta}{2\beta}  \right) 
-\ff \left( -\frac{\Delta_2-\beta}{2\beta} \right) 
\right. \no\\ &&  
 -\ff \left( 1+\frac { \left( \Delta_2-\beta \right) \left( \Delta_1-\beta \right) }{4\beta} \right) 
 +\ff \left( 1-\frac{ \left( \Delta_2+\beta \right)  \left( \Delta_1+\beta \right) }{4\beta}  \right) 
\no\\ &&  
-\ln  \left( \Delta_2-\beta \right) \ln  \left( \Delta_1+\beta \right) 
+\ln  \left( \Delta_2+\beta \right) \ln  \left( \Delta_1-\beta \right) 
+\ln  \left( 2 \right) \ln  \left( {\frac {\Delta_2-\beta}{\Delta_2+\beta}} \right) 
\no\\ &&  \left.
+\ln  \left( \beta \right) \ln  \left( {\frac {\Delta_1+\beta}{\Delta_1-\beta}} \right)  \right] 
+2 +3\,\ln  \left(  \frac{{\Delta_1}^{2}-{\beta}^{2}  }{4} \right)
\no \\ &&  
-\frac{1}{{\Delta_2}^{2}-{\beta}^{2} } \left[ 2\,\Delta_2\,
\ln \left( \frac{{\Delta_1}^{2}-{\beta}^{2}}{4} \right) 
+2\,\beta\,\ln  \left( {\frac {\Delta_1+\beta}{\Delta_1-\beta}} \right)  \right] 
\\ && \no  \\
N_{\pm 0} &=& \frac{3}{2} \ln\left(\frac{\mu^2}{M^2_H}\right) - \frac{3}{2} \ln(r_1^2)
\eea
In both cases ($H^0 \to P_1^+ P^-_2$ and $H^\pm \to P_1^\pm P^0_2$)
we fully recover the results of Ref.~\cite{Cir} in the limiting case 
$r_1=r_2$ and setting $\mu = M_1$.

\begin{figure}[t]
\begin{center}
\includegraphics[width=7.8cm]{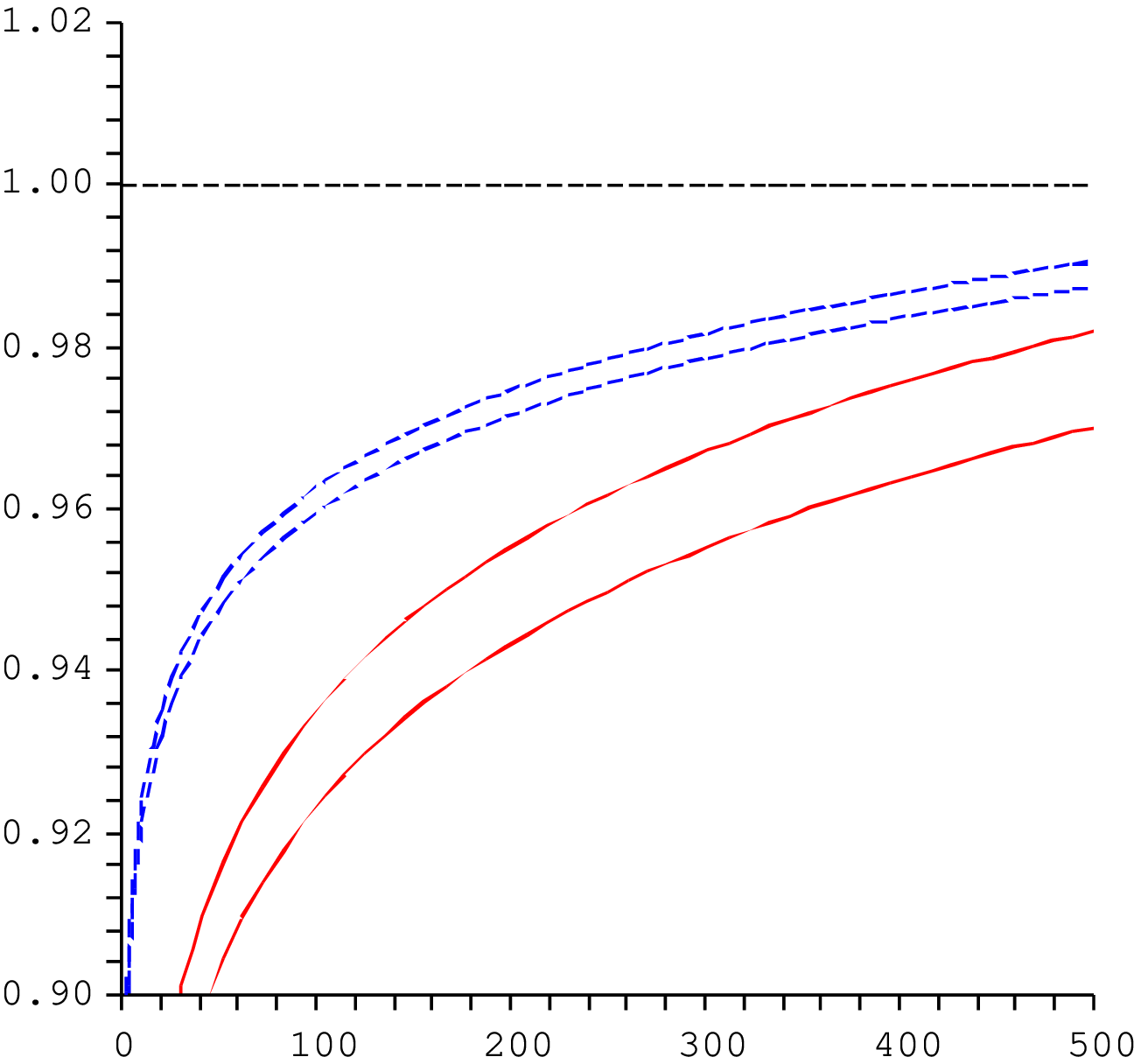}
\includegraphics[width=7.8cm]{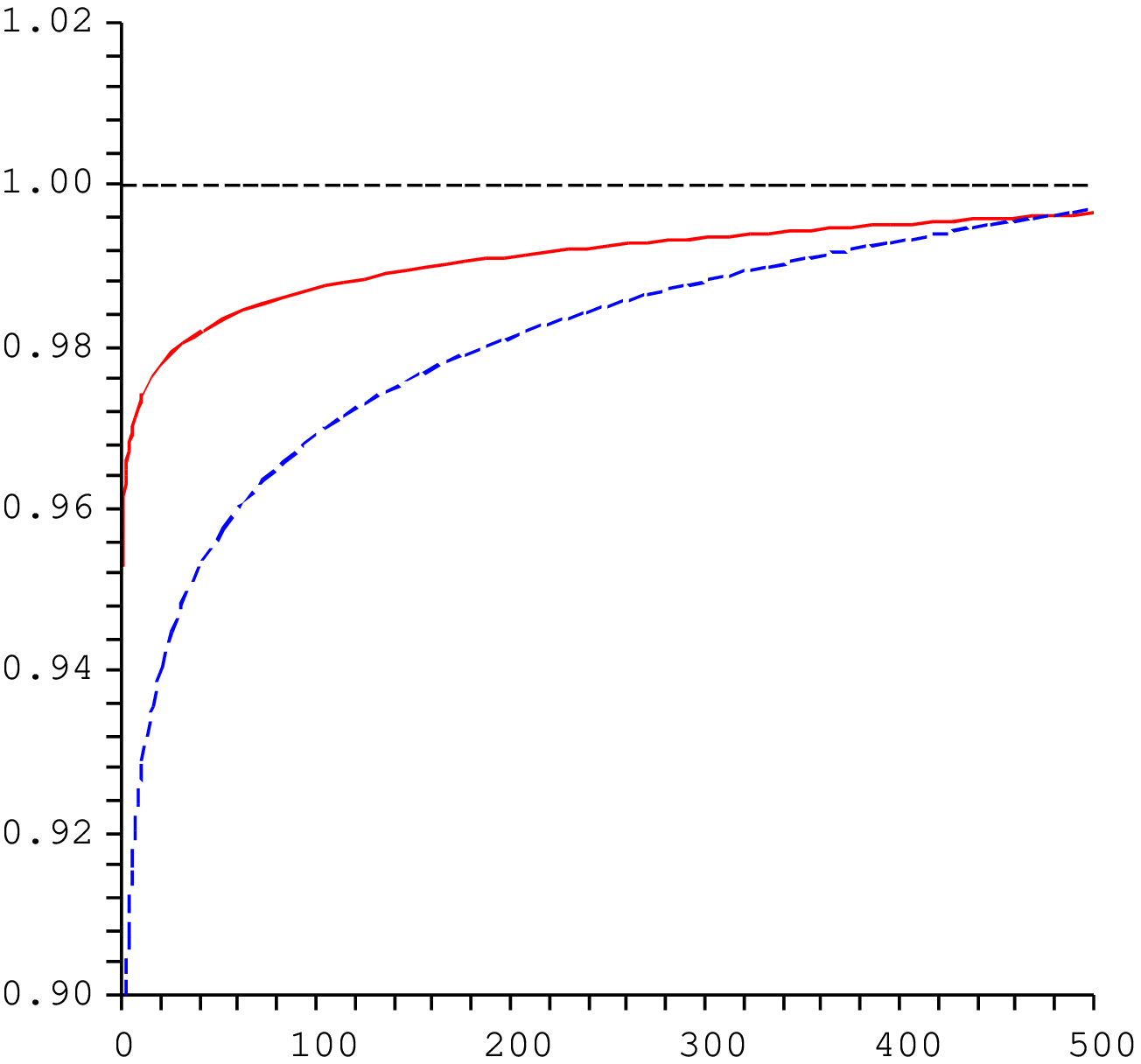}
\end{center}
\vspace{-20 pt}
\hspace{90 pt} 
\footnotesize{$E~(\rm MeV)$  }  
\hspace{176 pt} 
\footnotesize{$E~(\rm MeV)$  }
\caption{Examples of the~e.m.~correction factors for the photon-inclusive 
widths as a function of the cut on the maximal missing energy.
Left: $G_{\pi^+\pi^-}$ (full lines) and $G_{K^+K^-}$ (dashed lines)
for the corresponding $B$ decays; in each set the two curves are obtained for 
$\mu = M_\rho$ (upper curve) and $\mu = M_\pi (M_K)$ (lower curve).
Right: scale-independent ratios  $G_{K^0\pi^+}/G_{K^+\pi^0}$ 
(full line) and $G_{\pi^+\pi^-}/G_{\pi^+\pi^0}$ 
(dashed line) for the corresponding $B$ decays. \label{fig:plots} }
\end{figure}

\section{Numerical results and differential distributions}
In figure~\ref{fig:plots} we report a few examples 
of the~e.m.~correction factors relevant 
to $B \to \pi\pi$, $\pi K$, and $K K$ modes. 
As can be noted, the effects are quite large in case 
of tight cuts on the photon energy or, equivalently, 
tight cuts on the invariant mass of the two pseudoscalar
mesons. The e.m.~effects grow logarithmically with respect 
to the velocities of the charged particles in the final state
and with respect to the cut on the maximal photon energy (relative    
to the mass of the decaying particle). As a result, 
the largest effects are found for $B$ decays into charged pions. 
For instance, the {\em observable ratio} 
$\Gamma[B\to \pi^+\pi^-(\gamma)]/\Gamma[B\to\pi^0\pi^0]$,
determined with a cut of about 250 MeV on the missing 
mass of the $\pi^+\pi^-$ pair, receives a  $5\%$ negative 
isospin-breaking correction with respect to the corresponding 
{\em theoretical ratio} evaluated in absence of soft-photon 
contributions.

As anticipated, the overall normalization of the $G_{12}(E)$ factors 
is convention dep\-en\-dent. By construction, they include the whole 
effect of the bremsstrahlung but only part of the virtual e.m.~corrections, 
namely the leading soft contribution which can be reliably estimated 
within scalar QED. 
The separation between hard and soft virtual contributions 
is arbitrary and determined by the choice of the UV renormalization 
scale~$\mu$. The most natural choice for the latter is between 
the light pseudoscalar masses ($M_\pi$ or $M_K$)
and $M_\rho \approx$~770~MeV. As shown in the left plot of 
figure~\ref{fig:plots}, the uncertainty associated with 
this range is at most of $\cO(1\%)$ and definitely subleading 
with respect to the leading effects induced by soft contributions.

In principle, this scale uncertainty could be removed by means 
of an appropriate matching between our effective theory --where photons and
mesons are the only dynamical degrees of freedom and each weak amplitude 
is determined by a momentum-independent coupling-- 
and more predictive effective theories where the weak decay 
amplitudes are computed in terms of the fundamental SM couplings. 
In the last few years there has been a substantial progress 
toward a precise estimate of non-leptonic weak decay amplitudes within 
the SM (see e.g.~Ref.~\cite{Vari-Bpp} and references therein); however, 
we are still far from the $\cO(1\%)$ level, especially for the decays 
into two light pseudoscalar mesons. Moreover, all present approaches 
do not include the structure-dependent electromagnetic corrections which 
would allow to match the scale dependence of the $G_{12}(E)$ factors.

Despite the absence of a precise UV matching, in a few cases
the e.m.~correction factors we have computed allows to evaluate in 
a precise way the amount of isospin breaking induced by soft photons. 
As illustrated in the right plot of figure~\ref{fig:plots},
the scale uncertainty drops out completely 
in all the ratios of two-body photon-inclusive widths  
with at least one charged particle in the final state.
    
It is worth to stress that the e.m.~corrections discussed 
in this work can be one of the sources of the ``anomalous'' 
isospin-breaking effects identified 
in recent phenomenological analyses of $B\to \pi K$ and $B\to \pi\pi$ 
decays \cite{Flei}.  
A consistent phenomenological analysis of these channels with 
the inclusion of radiative corrections requires the information 
on the experimental cuts applied on the soft-photon radiation
and is beyond the purpose of the present work. However, it is 
clear that  radiative corrections
 need to be included in these channels,
especially if one is interested in isospin-breaking effects 
as a tool to identify possible deviations from the SM.

\medskip

From the purely experimental side, the only relevant 
aspect of the $G_{12}(E)$ factors is their energy dependence,
which is unambiguously determined up to $\cO(E)$ terms.
This allows us to evaluate the observable missing-energy distribution,
or the soft-photon spectrum, $d \Gamma^{\rm incl}_{12}(E)/d E$.
As is well known, the $E\to 0$ singularity of this distribution 
is not integrable if evaluated at any fixed order in perturbation
theory; however, the all-order  resummation of the leading infrared 
singularities leads to an integrable 
distribution \cite{Yen,Wei}. In our case, this can be written as
\be
\frac{d \Gamma^{\rm incl}_{12}(E)}{dE} 
~=~ \frac{2 \alpha}{\pi}  ~\frac{|b_{12}| \Gamma^0_{12} }{E}
\left( \frac{2E}{M_H} \right)^{ \frac{2 \alpha }{\pi}  |b_{12}| } 
\left[ 1+\cO\left( \frac{E}{M_H}, \frac{\alpha}{\pi}\right) \right]
\ee
with the $b_{12}$ coefficients given in \eq{b12}. The exponentiation of the 
leading-log corrections does not lead to appreciable numerical 
difference with respect to the pure $\cO(\alpha)$ result for missing 
energies above few MeV. However, as discussed in \cite{Gatti},
having an integrable distribution can be very useful in preparing   
a dedicated Montecarlo program.

Concerning the angular distribution of the bremsstrahlung photons, 
the $\cO(\alpha)$ result reads
\be
\frac{d^2 \Gamma(H \to P_1 P_2\gamma) }{ d E_\gamma ~d \cos\theta_\gamma } =
\frac{\alpha}{2\pi}~\frac{\beta_\gamma}{\beta}~\frac{\Gamma^0_{12}}{E_\gamma }~R_{12} 
\ee
where $E_\gamma$  and $\theta_\gamma$ denote, respectively, 
photon energy and angle between 
photon and $P_1$ momenta in the $H$ meson rest frame, and 
\be
\beta^2 ~=~ \left[ 1- \frac{(r_1+r_2)^2}{1-2z} \right] \left[ 1- \frac{(r_1-r_2)^2}{1-2z}\right]~, 
\qquad z~=~ \frac{E_\gamma}{M_H}~.
\ee
The $R_{12}$ coefficients, defined by 
\be
R_{12} ~=~  E_\gamma^2 \sum_{\rm spins} 
\left| \frac{\cA [H(p_H) \to P_1(p_1) P_2(p_2)\gamma]}{e \cA  (H \to P_1 P_2)} \right|^2~=~ 
E_\gamma^2 \left| \sum_{i=H,P1,P2} Q_i \frac{p_i^\mu}{k\cdot p_i} \right|^2~,
\ee
where $Q_{1,2}$ and $-Q_H$ denote, respectively,  
the electric charges of $P_{1,2}$ and $H$ in units of the 
electron charge, assume the following explicit form 
\bea
R_{+-} &=& \frac{1-r_1^2 -r_2^2 -2z}{t_1 t_2} - \frac{r_1^2}{t_1^2}- \frac{r_2^2}{t_2^2}~, \\
R_{\pm 0} &=& \frac{1+r_1^2 -r_2^2 -2z (1-t_1)}{t_1} - \frac{r_1^2}{t_1^2} - 1~, 
\eea
in terms of the kinematical variables
\be
t_{1,2} = \frac{1}{2} 
\left[ 1 + \frac{r_{1,2}^2}{1-2z} \mp \beta_\gamma \cos\theta_\gamma \right]~.
\ee

\section{Conclusions}
In the last few years there has been a substantial progress
in the experimental determination of two-body non-leptonic 
$B$ decays into light pseudoscalar mesons. 
Several new theoretical tools have also been developed 
for the description of these interesting processes. 
However, in most cases electromagnetic effects 
of long-distance origin have been ignored,  
both in the theoretical predictions and in the experimental analyses.
Aiming at accuracies of the order of a few percent,
this approximation is no longer valid.

In this work we have presented a detailed 
discussion of the electromagnetic effects  
of long-distance origin in two-body non-leptonic 
$B$ decays. 
In particular, we have computed the leading $\cO(\alpha)$ effects 
induced by both real and virtual photons in a generic process 
of the type $H \to P_1 P_2(\gamma)$, where both $H$ and $P_{1,2}$
are scalar or pseudoscalar particles (with arbitrary masses).
The results thus obtained can be applied to both $B$ and $D$ decays. 
 
We have performed the calculation within scalar QED and assuming 
an effective constant coupling for the weak vertex. 
Given the universal character of the infrared singularities, 
this simplified framework allows to identify the leading 
effects induced by soft photons.  The results obtained 
within this framework are valid up to energy-independent 
$\cO(\alpha)$ terms related to the matching between this effective theory and the SM, 
and possible $\cO(E_\gamma)$ terms associated to the emission of 
hard photons (which should be identified experimentally). 
In the case of the photon-inclusive rates, the analytic formulae 
we have obtained represent a generalization of the results of
Ref.~\cite{Cir}, which are recovered in the limiting case $M_1=M_2$. 
In order to facilitate the experimental simulation of the 
soft-photon radiation, we have also discussed the corresponding 
angular and energy distribution. 

From the phenomenological point of view, the largest impact 
of the soft-photon radiation is found in the $B \to \pi^+ \pi^-$ channel,
where the corrections to the partially-inclusive rate can easily 
exceed the $5\%$ level. We stress that detailed estimate of these 
effects can only be made once the experimental 
information on the photon cuts (or the missing energy) is 
available. Without this information, an unambiguous comparison 
between theory and experiments, and also the combination of 
different experimental results, cannot be performed.

\section*{Acknowledgments}
We thank G.~Cavoto, V.~Cirigliano, F.~Ferroni, G.~Martinelli, M.~Pierini, and L.~Silvestrini 
for useful discussions. This work was supported in part by the IHP-RTN program, 
EC contract No.~HPRN-CT-2002-00311 (EURIDICE).

\end{document}